\documentclass[10pt,conference,letterpaper]{IEEEtran}
\IEEEoverridecommandlockouts
\pdfoutput=1

\usepackage[T1]{fontenc}
\usepackage{cite}
\usepackage{amsmath,amssymb,amsfonts}
\usepackage{amsthm}
\usepackage{float} 
\usepackage{enumitem}
\usepackage{algorithm}
\usepackage[noend]{algpseudocode}
\usepackage{makecell}

\usepackage{stfloats}
\usepackage{graphicx}
\usepackage{textcomp}
\usepackage{dcolumn} 
\usepackage{bm}      
\usepackage{braket}
\usepackage{tikz}
\usetikzlibrary{quantikz}
\usetikzlibrary{positioning,fit,calc}
\usepackage{hyperref} 
\hypersetup{
    colorlinks = true,
    citecolor = magenta,
    linkcolor = purple
}

\newtheorem{definition}{Definition}

\algrenewcommand\algorithmicrequire{\textbf{Input:}}
\algrenewcommand\algorithmicensure{\textbf{Output:}}

\def\BibTeX{{\rm B\kern-.05em{\sc i\kern-.025em b}\kern-.08em
    T\kern-.1667em\lower.7ex\hbox{E}\kern-.125emX}}

\begin{document}
\bstctlcite{IEEEexample:BSTcontrol}

\title{Quantum Circuit Design for Decoded Quantum Interferometry
\thanks{
This work was supported by MEXT-Quantum Leap Flagship Program Grant
Number JPMXS0118067285 and JPMXS0120319794. JST Moonshot R$\&$D Grant Numbers JPMJMS2061 and JPMJMS226C.}
}

\author{
\IEEEauthorblockN{
Natchapol Patamawisut\IEEEauthorrefmark{1}\IEEEauthorrefmark{3},
Naphan Benchasattabuse\IEEEauthorrefmark{1}\IEEEauthorrefmark{3},
Michal Hajdu\v{s}ek\IEEEauthorrefmark{1}\IEEEauthorrefmark{3},
and Rodney Van Meter\IEEEauthorrefmark{2}\IEEEauthorrefmark{3}}\\

\IEEEauthorblockA{\IEEEauthorrefmark{1}\textit{Graduate School of Media and Governance, Keio University Shonan Fujisawa Campus, Kanagawa, Japan}}
\IEEEauthorblockA{\IEEEauthorrefmark{2}\textit{Faculty of Environment and Information Studies, Keio University Shonan Fujisawa Campus, Kanagawa, Japan}}
\IEEEauthorblockA{\IEEEauthorrefmark{3}\textit{Quantum Computing Center, Keio University, Kanagawa, Japan}\\
\{banknatchapol, whit3z, michal, rdv\}@sfc.wide.ad.jp}
}

\maketitle

\begin{abstract}
Decoded Quantum Interferometry (DQI) is a recently proposed quantum algorithm for approximating solutions to combinatorial optimization problems by reducing instances of linear satisfiability to bounded-distance decoding over superpositions of quantum states.
A central challenge in realizing DQI is the design of a decoder that operates coherently on quantum superpositions.
In this work, we present a concrete quantum circuit implementation of DQI, with a focus on the decoding subroutine.
Our design leverages a reversible Gauss-Jordan elimination circuit for the decoding stage.
We analyze the circuit’s depth and gate complexity and validate its performance through simulations on systems with up to 30 qubits.
These results establish a concrete foundation for scalable implementations of DQI and open the door to future algorithmic refinements and hardware-level realizations.
\end{abstract}

\section{Introduction}
Quantum computing promises significant advantages over classical computing in domains such as scientific simulations, machine learning, and combinatorial optimization~\cite{ref15}\cite{Abbas_2021}\cite{bärtschi2025potentialapplicationsquantumcomputing}. Leading approaches like quantum annealing~\cite{Quinton2025} and the Quantum Approximate Optimization Algorithm (QAOA)~\cite{ref8} encode optimization problems into Hamiltonians to find low-energy (near-ground-state) solutions~\cite{ref9}.

Despite promising numerical results for specific problems~\cite{ref17}~\cite{ruslan-labs-evidence-of-qaoa-scaling}~\cite{john-numerical-speedup-constrained}~\cite{john-sat-qaoa}, iterative versions of Grover’s algorithm~\cite{ref4}~\cite{gilliam-grover-adaptive-search-cpbo}~\cite{durrQuantumAlgorithmFinding1999} and its variants~\cite{koch-variational-amplitude-amplification}~\cite{ref16}—designed as quantum brute-force searches—offer only a quadratic speedup. This gain is generally insufficient for practical advantage, especially given the overhead required for fault-tolerant error correction~\cite{Fowler_2012}. Thus, to achieve meaningful quantum advantage, algorithms must exploit inherent structural properties of optimization problems.

Motivated by this need, Jordan et al.~\cite{jordan-dqi} recently proposed Decoded Quantum Interferometry (DQI), which leverages quantum interference to amplify state amplitudes in proportion to their objective values. Their analysis shows that certain combinatorial problems (e.g., maximum linear satisfiability) exhibit sparsity in their Fourier spectrum—a feature that DQI explicitly exploits. This interferometric decoding paradigm offers a distinct alternative to Hamiltonian-based methods and holds potential for exponential speedups in select classes.

However, while the original proposal provided high-level circuit designs, the critical decoding subroutine was only tested classically using belief propagation techniques. No explicit, fully reversible quantum circuit for decoding was provided, leaving a key gap for practical deployment.

In this work, we present explicit quantum circuit constructions for all components of the DQI algorithm, tailored to the Max-LINSAT problem over the binary field (also known as Max-XORSAT). Notably, we demonstrate that the decoding subroutine can be implemented using reversible Gauss–Jordan elimination, constructed entirely from CNOT and Toffoli gates~\cite{ref3}. We conduct a comprehensive resource analysis—covering gate counts, circuit depths, and qubit requirements—and validate our designs with simulations on problem instances involving up to 30 qubits for MaxCut problems (an instance of Max-XORSAT).

\section{Decoded Quantum Interferometry}
There are three common ways to encode a classical function \( f(x) \) into a quantum state:

\begin{enumerate}
    \item Entangled encoding: Prepare a joint state over two registers such that one holds the input and the other holds the corresponding output:
    \begin{equation}
    \ket{\psi} = \sum_x \ket{x} \ket{f(x)}.
    \end{equation}
    \item Phase encoding: Encode the function output in the phase of each basis state:
    \begin{equation}
    \ket{\psi} = \sum_x e^{-i f(x)} \ket{x}.
    \end{equation}
    \item Amplitude encoding: Encode the function values directly as amplitudes:
        \begin{equation}
    \ket{\psi} = \sum_x f(x) \ket{x}.
    \end{equation}
\end{enumerate}
In each case, we omit normalization constants for simplicity, but the resulting quantum state must be properly normalized to ensure a valid quantum state.

Amplitude encoding is especially powerful for optimization, since if one can prepare such a state, measuring it yields bitstrings with probabilities biased toward higher objective values. However, preparing amplitude-encoded states for unstructured functions is challenging. A common workaround~\cite{Biamonte_2017} is to assume access to a Quantum Random Access Memory (QRAM)~\cite{Giovannetti_2008} for efficient superposition-based data retrieval.

Jordan et al.~\cite{jordan-dqi} observed that, for instances of Max-LINSAT, the amplitude-encoded state is sparse in the Hadamard basis. In other words, applying a Hadamard transform to the amplitude-encoded state results in a state with support only on a small subset of basis states. This observation permits efficient preparation of the Hadamard-transformed state, which can then be inverted to obtain the desired amplitude encoding. This insight is central to the DQI framework for biasing quantum interference toward optimal or near-optimal solutions.

Decoded Quantum Interferometry (DQI) leverages the sparsity of the Fourier spectrum of the objective function to amplify amplitudes corresponding to optimal solutions via quantum interference. For the Max‑XORSAT problem over \(\mathbb{F}_2\), one is given a binary matrix \(B \in \{0,1\}^{m \times n}\) (representing \(m\) constraints on \(n\) variables) and a vector \(v \in \mathbb{F}_2^m\). The goal is to find an \(n\)-bit string \(x\) that maximizes 
\begin{equation}
f(x) = \sum_{i=1}^{m} (-1)^{v_i + b_i \cdot x},
\label{eq:fx}
\end{equation}
where \(b_i\) is the \(i\)th row of \(B\). Since \(f(x)\) represents the difference between the number of satisfied and unsatisfied equations, its Hadamard (Fourier) transform is sparse; in fact,
\begin{equation}
\sum_{x\in\mathbb{F}_2^n} f(x)|x\rangle \propto \sum_{i=1}^{m} (-1)^{v_i} |b_i\rangle.
\end{equation}
This sparsity indicates that the state can be biased toward solutions with larger values of $f(x)$. However, because $f(x)$ may be negative—and we do not wish to amplify those values—we introduce a polynomial $P(f(x))$ of degree $\ell$, chosen to match the maximum number of bit-flip errors the classical decoder can correct, to selectively amplify only near-optimal (i.e., high) outputs. In this way, the state
\begin{equation}
|P(f)\rangle = \sum_{x\in\mathbb{F}_2^n} P\bigl(f(x)\bigr)\,\lvert x\rangle,
\end{equation}
can be efficiently prepared to favor measurement outcomes corresponding to higher values of $f(x)$.
This sparsity indicates that the state can be biased toward solutions with larger values of $f(x)$. However, because $f(x)$ may be negative—and we do not wish to amplify those values—we introduce a polynomial $P(f(x))$ of degree $\ell$, chosen to match the maximum number of bit-flip errors the classical decoder can correct, to selectively amplify only near-optimal (i.e., high) outputs. In this way, the state
\begin{equation}
|P(f)\rangle = \sum_{x\in\mathbb{F}_2^n} P(f(x)) |x\rangle,
\label{eq:Pf}
\end{equation}
can be efficiently prepared to favor measurement outcomes corresponding to higher values of \(f(x)\).

A key result (Theorem 4.1 in~\cite{jordan-dqi}) relates the expected number of satisfied constraints \(\langle s \rangle\) to the parameters \(r\) (the number of inputs yielding \(+1\)), \(p\) (the prime defining \(\mathbb{F}_p\)), and \(\ell\):
\begin{equation}
\frac{\langle s\rangle}{m} \;=\;
\left(\sqrt{\frac{\ell}{m}\Bigl(1 - \frac{r}{p}\Bigr)} +
\sqrt{\frac{r}{p}\Bigl(1 - \frac{\ell}{m}\Bigr)}\right)^2
\quad \text{if } \frac{r}{p} \leq 1 - \frac{\ell}{m},
\end{equation}
with \(\langle s\rangle/m = 1\) otherwise.

\subsection*{DQI for Max-XORSAT: Quantum State Evolution}

The quantum-state evolution of DQI proceeds in seven stages:

\subsubsection{Amplitude Encoding}  
Starting from \(\ket{0}^{\otimes m}\), we apply a unitary to encode the weight vector \(\{w_k\}_{k=0}^{\ell}\) in the state
\[
\sum_{k=0}^{\ell} w_k\, \ket{k},
\]
where the \(w_k\) are the optimal coefficients obtained from the principal eigenvector of an \((\ell+1)\times(\ell+1)\) matrix (see Lemma 9.2 in~\cite{jordan-dqi}). We call this the \emph{error register}, as it eventually encodes potential error information.

\subsubsection{Dicke State Preparation}  
The amplitude-encoded state is transformed into a Dicke state:
\begin{equation}
\sum_{k=0}^{\ell} w_k \,|D(m,k)\rangle,
\end{equation}
with
\begin{equation}
\label{eq:DickeState}
|D(m,k)\rangle = \frac{1}{\sqrt{\binom{m}{k}}}\sum_{\substack{y \in \mathbb{F}_2^m \\ |y|=k}} |y\rangle.
\end{equation}

\subsubsection{Phase Encoding}  
Each qubit of the error register is phase shifted according to
\[
\ket{y} \mapsto (-1)^{v \cdot y}\ket{y}.
\]

\subsubsection{Constraint Encoding}  
Conditional operations embed the binary matrix \(B\) into the syndrome register, constructing the state
\[
\ket{B^T y}.
\]
Here, \(\ket{y}\) in the error register represents the error and \(B^T\) acts as the parity check matrix, so the syndrome register holds the error syndrome.

\subsubsection{Decoding}  
A unitary decoding procedure converts \(\ket{B^T y}\) back to \(\ket{y}\) and then uses controlled XOR (CNOT) operations to uncompute the error register, thereby isolating the syndrome information.

\subsubsection{Quantum Fourier Transform}  
For Max-XORSAT, applying Hadamard transforms to the syndrome register yields
\[
\sum_{x \in \{0,1\}^n} P\bigl(f(x)\bigr)\,\ket{x},
\]
biasing measurements toward higher objective values.

\subsubsection{Measurement and Postselection}  
Finally, measuring all registers and post-selecting on the error register being \(\ket{0}^{\otimes m}\) produces the approximate solution \(x\).

\section{Dicke State Preparation}
\label{sec:dicke}

Dicke states are entangled states with a fixed number of excitations (i.e., qubits in the \(|1\rangle\) state). For an \(n\)-qubit system with \(h\) excitations, the Dicke state is defined as
\begin{equation}
\label{eq:DickeState}
|D(n,h)\rangle 
=\frac{1}{\sqrt{\binom{n}{h}}} 
\sum_{\substack{y \in \{0,1\}^n \\ |y| = h}} |y\rangle,
\end{equation}
which is symmetric under any qubit permutation. Dicke states have attracted attention for their role in quantum metrology~\cite{ref18}, quantum optimization~\cite{ref19}, and other applications.

\vspace{0.75em}
\noindent
\textbf{Deterministic Preparation Protocol.}
Instead of relying on probabilistic postselection~\cite{841192}, Bärtschi and Eidenbenz~\cite{ref2} propose a deterministic algorithm that builds \(\ket{D(n,h)}\) from smaller states by defining a family of unitaries \(U_{n,h}\) that map partially prepared Dicke states to larger ones.

\begin{definition}[Dicke-State Unitary \(U_{n,h}\)]
\label{def:Unh}
For \(0 \le h \le n\), \(U_{n,h}\) is a unitary acting on \(n\) qubits that transforms the basis state \(\ket{0}^{\otimes n-s} \ket{1}^{\otimes s}\) into \(\ket{D(n,s)}\) for all excitations \(s \le h\).
\end{definition}

\subsection{Split and Cyclic Shift (SCS) Unitary}
A key component is the \emph{Split and Cyclic Shift} (\(\mathrm{SCS}\)) unitary, \(\mathrm{SCS}_{n,h}\), which redistributes amplitudes according to the Dicke-state recursion:
\[
\sqrt{\frac{n-h}{n}}\,\ket{0}\otimes \ket{D(n-1,h)}
+\sqrt{\frac{h}{n}}\,\ket{1}\otimes \ket{D(n-1,h-1)}.
\]
In other words, \(\mathrm{SCS}_{n,h}\) "splits" the amplitude based on whether the new qubit is \(\ket{0}\) or \(\ket{1}\), and "cycles" the states as needed.

\begin{definition}[Split and Cyclic Shift (SCS)]
\label{def:SCS_gate}
For \(0 \le h < n\), a unitary \(\mathrm{SCS}_{n,h}\) acts on (at least) \(n\) qubits and redistributes amplitude from the Dicke state \(\ket{D(n-1,h)}\) to the states
\[
\ket{0}\otimes \ket{D(n-1,h)} \quad \text{and} \quad \ket{1}\otimes \ket{D(n-1,h-1)},
\]
in accordance with the recursion above.
\end{definition}

Bärtschi and Eidenbenz~\cite{ref2} further decompose \(\mathrm{SCS}\) into two elementary operations: the two-qubit gate, \(\mathrm{SCS}_2\), and the three-qubit gate, \(\mathrm{SCS}_3\) (corresponding to building blocks (i) and (ii), respectively).

\subsection{Construction of \(U_{n,h}\) from SCS}
\label{subsec:inductive-construction}

The unitary \(U_{n,h}\) can be built inductively from \(U_{n-1,h}\) by applying a series of SCS unitaries to the newly added qubit along with the first \(n-1\) qubits. Formally, we have:
\begin{equation}
\label{eq:Unh_product}
\begin{split}
U_{n,h} 
\;=\;
\prod_{s=2}^{h}
\Bigl(\mathrm{SCS}_{s,s-1} \,\otimes\, \mathrm{Id}^{\otimes(n-s)}\Bigr) \\& 
\hspace{-4.56cm}\cdot \prod_{s=h+1}^{n}
\Bigl(\mathrm{Id}^{\otimes(s-h-1)} \,\otimes\,\mathrm{SCS}_{s,h} \,\otimes\, \mathrm{Id}^{\otimes(n-s)}\Bigr).
\end{split}
\end{equation}
Here, each \(\mathrm{SCS}_{n,h}\) acts on \(n+1\) qubits (or \(n\) plus one ancilla) to perform the amplitude-splitting, while the identity operators ensure proper application across the entire register.

\section{Quantum Circuit Implementation of Gauss-Jordan Elimination}
Many problems involving the solution of linear systems or the decoding of classical information—such as the binary syndrome decoding problem~\cite{ref12}—can benefit from implementing Gauss–Jordan elimination reversibly. In quantum computing, this requires performing row operations that lead to the reduced row-echelon form while preserving quantum coherence. In the Information Set Decoding (ISD) algorithm—central to attacking code-based cryptosystems like Classic McEliece~\cite{ref20} and BIKE~\cite{ref21}—Gaussian elimination is used to invert a submatrix of the parity-check matrix, transforming the syndrome vector and identifying low-weight error vectors. Reversible Gauss–Jordan elimination thus enables efficient quantum implementation of ISD~\cite{ref3}.

Several works, including Perriello et al.~\cite{ref3} and optimized implementations~\cite{ref6}\cite{ref7}, demonstrate that Gaussian elimination can be carried out using reversible operations primarily composed of CNOT and SWAP gates.

The basic idea is to map the standard Gaussian elimination procedure onto a quantum circuit acting on an augmented matrix \(A\) (encoding \([B^T\,|\,s]\)):
\begin{enumerate}
    \item For each column \( j \) in \( [B^T\,|\,s] \):
    \begin{enumerate}
        \item Pivot Selection: Identify a pivot row \( i \) with \( A_{i,j}=1 \).
        \item Row Swapping: If \( i \neq j \), swap row \( j \) with row \( i \) using quantum SWAP gates.
        \item Row Elimination: For every other row \( k \neq j \) with \( A_{k,j}=1 \), apply a CNOT gate (control on row \( j \), target on row \( k \)) to clear the 1 in column \( j \).
    \end{enumerate}
    \item Repeat for all columns. The reversibility ensures that, once \(A\) is in reduced row-echelon form, the state corresponding to the error vector \( y \) is uncomputed.
\end{enumerate}
The precise steps of this algorithm are shown in Algorithm~\ref{alg:gaussjordan}.

\begin{algorithm}[tb]
\caption{Quantum Circuit Friendly Gauss-Jordan Elimination}
\label{alg:gaussjordan}
\begin{algorithmic}[1]
\Require Augmented matrix register \(A\) representing \([B^T\,|\, s]\).
\Ensure Reduced row-echelon form of \(A\) and extraction of \(y\).
\For{each column \(j = 1\) to \(n\)}
    \State Find a pivot row \(i\) such that \(A_{i,j} = 1\);
    \If{\(i \neq j\)}
        \State Swap row \(j\) and row \(i\) using quantum SWAP gates.
    \EndIf
    \For{each row \(k \neq j\)}
        \If{\(A_{k,j} = 1\)}
            \State Apply CNOT with control on row \(j\) and target on row \(k\).
        \EndIf
    \EndFor
\EndFor
\State \textbf{Output:} \(A\) is in reduced row-echelon form and encodes \(y\).
\end{algorithmic}
\end{algorithm}

\section{Implementation Approach}
The DQI algorithm decodes an error pattern by preparing specific quantum states and applying a series of transformations. Our implementation proceeds as follows:


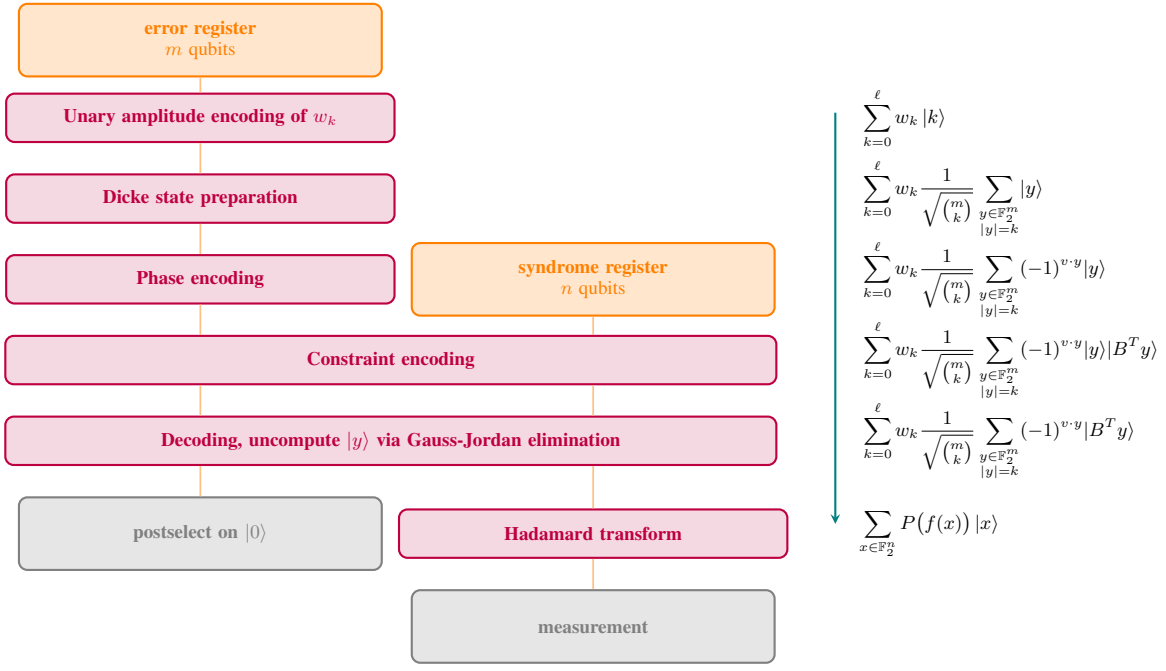
\begin{figure*}[htbp]
\centering
\scalebox{0.8}{
\begin{tikzpicture}[
  font=\small,
  Orange/.style={
    draw=orange,
    line width=1.0pt,
    fill=orange!20,
    rounded corners,
    align=center,
    minimum height=1.2cm,
    text width=3cm,
    inner sep=6pt,
    text=orange,
    minimum width=6cm,
  },
  Red/.style={
    draw=purple,
    line width=1.0pt,
    fill=purple!15,
    rounded corners,
    align=center,
    minimum height=.8cm,
    text width=6cm,
    inner sep=6pt,
    text=purple,
    minimum width=6cm,
  },
  tealBox/.style={
    draw=teal,
    line width=1pt,
    fill=teal!10,
    rounded corners,
    align=center,
    minimum height=1.2cm,
    text width=3cm,
    inner sep=6pt,
    text=teal,
    minimum width=6cm,
  },
  Gray/.style={
    draw=gray,
    line width=1.0pt,
    fill=gray!20,
    rounded corners,
    align=center,
    minimum height=1.2cm,
    text width=3cm,
    inner sep=6pt,
    text=gray,
    minimum width=6cm,
  },
  arrowstyle/.style={
    thick,
    draw=orange!50
  }
]

\node[Orange] (errorreg) {%
  \textbf{error register}\\
  \( m \) qubits
};

\node[Red, below=.25cm of errorreg] (embed) {%
  \textbf{Unary amplitude encoding of \( w_k \)}
};

\node[below=.25cm of embed, right=7.61cm of embed] (state1) {
  $\displaystyle \sum^{\ell}_{k=0} w_{k}\,| k\rangle$
};

\node[Red, below=.5cm of embed] (dicke) {%
  \textbf{Dicke state preparation}
};

\node[below=.25cm of dicke, right=7.61cm of dicke] (state2) {
 $\displaystyle 
 \sum_{k=0}^{\ell} 
w_k \frac{1}{\sqrt{\binom{m}{k}}}
\sum_{\substack{y \in \mathbb{F}_2^m \\ |y| = k}}
\lvert y\rangle
 $
};

\node[Red, below=.5cm of dicke] (z) {%
  \textbf{Phase encoding}
};

\node[below=.25cm of z, right=8.01cm of z] (state3) {
 \hspace*{-.4cm}$\displaystyle 
 \sum_{k=0}^{\ell} 
w_k \frac{1}{\sqrt{\binom{m}{k}}}
\sum_{\substack{y \in \mathbb{F}_2^m \\ |y| = k}}
(-1)^{v \cdot y}\lvert y\rangle
 $
};

\node[Orange, right=.25cm of z] (syndromereg) {%
  \textbf{syndrome register}\\
  \( n \) qubits
};

\node[Red, below=.5cm of z, xshift=3.15cm, minimum width=12.75cm] (BT) {%
  \textbf{Constraint encoding}
};

\node[right=1.7cm of BT] (state4) {
 \hspace*{-.4cm}$\displaystyle 
 \sum_{k=0}^{\ell} 
w_k \frac{1}{\sqrt{\binom{m}{k}}}
\sum_{\substack{y \in \mathbb{F}_2^m \\ |y| = k}}
(-1)^{v \cdot y}\lvert y\rangle \lvert B^{T}y\rangle
 $
};

\node[Red, below=.5cm of BT, minimum width=12.75cm, text width=12cm] (uny) {%
  \textbf{Decoding, uncompute \( \lvert y \rangle \) via Gauss-Jordan elimination}
};

\node[right=2cm of uny, align=left] (state5) {
 \hspace*{-.7cm}$\displaystyle 
 \sum_{k=0}^{\ell} 
w_k \frac{1}{\sqrt{\binom{m}{k}}}
\sum_{\substack{y \in \mathbb{F}_2^m \\ |y| = k}}
(-1)^{v \cdot y} \lvert B^{T}y\rangle
 $
};

\node[Gray, below=.5cm of uny, xshift=-3.15cm] (grayNode) {%
  \textbf{postselect on \( \lvert 0 \rangle \)}
};

\node[Red, right=.25cm of grayNode] (hadamard) {%
  \textbf{Hadamard transform}
};

\node[right=1.045cm of hadamard] (state6) {
 \hspace*{-.cm}$\displaystyle 
 \sum_{x \in \mathbb{F}_2^n} P\bigl(f(x)\bigr)\,\lvert x\rangle
 $
};

\node[Gray, below=.5cm of hadamard] (measure) {%
  \textbf{measurement}
};

\draw[arrowstyle] (errorreg) -- (embed);
\draw[arrowstyle] (embed) -- (dicke);
\draw[arrowstyle] (dicke) -- (z);
\draw[arrowstyle] (z.south) -- ($(z.south |- BT.north)$);
\draw[arrowstyle] (syndromereg.south) -- ($(syndromereg.south |- BT.north)$);

\draw[arrowstyle] ($(z.south |- BT.south)$) -- ($(z.south |- uny.north)$);
\draw[arrowstyle] ($(syndromereg.south |- BT.south)$) -- ($(syndromereg.south |- uny.north)$);
\draw[arrowstyle] ($(z.south |- uny.south)$) -- ($(z.south |- grayNode.north)$);
\draw[arrowstyle] ($(syndromereg.south |- uny.south)$) -- ($(syndromereg.south |- hadamard.north)$);
\draw[arrowstyle] ($(syndromereg.south |- hadamard.south)$) -- ($(syndromereg.south |- measure.north)$);

\draw [teal, -stealth,line width=1.0pt](10.5,-1.2) -- (10.5,-8);

\end{tikzpicture}
}
\caption{DQI algorithm state evolution diagram.}
\label{fig:dqidigram}
\end{figure*}

\begin{enumerate}
    \item Amplitude Encoding: We first initialize the error registers using Unary Amplitude Encoding (UAE) to obtain the \(\bm{w_k}\) state encoded in a unary \( \bm{k}\) representation. In Jordan et al.~\cite{jordan-dqi}, it is suggested that one could use a technique from Low, Kliuchnikov, and Schaeffer~\cite{ref22} for this purpose. However, since we use the deterministic Dicke state preparation procedure from Bärtschi and Eidenbenz~\cite{ref2}, we employ quantum fan-out~\cite{ref11} to encode the state in a unary representation, which requires only \(O(\ell)\) resources.

   \item Dicke State Preparation: With the unary state already prepared, we apply the unitary \(U_{m,\ell}\)~\cite{ref2} to transform it into a Dicke state. This approach not only simplifies the overall state preparation process but also provides a robust baseline for further improvements.

    \item Phase Encoding: Pauli-\(Z\) gates are applied according to the vector \(v\) to encode phase information.
    \item Constraint Encoding: Controlled operations use the binary matrix \(B\) to condition the error state \(y\) so that the syndrome register is set to \( |B^{T}y \rangle \).

    \item Decoding: Syndrome decoding is performed on the syndrome register using a reversible Gauss--Jordan elimination algorithm. This method translates the classical Gauss--Jordan procedure into a quantum circuit by replacing classical row swapping and XOR operations with SWAP and CNOT gates, respectively. We chose this approach because its direct correspondence to the classical algorithm makes the implementation straightforward and easy to follow. The procedure effectively uncomputes the error register state, isolating the correctly decoded syndrome.

    \item Hadamard Transform: After uncomputation, a Hadamard transform is applied to the syndrome register to complete the transformation into the measurement basis.
    \item Measurement and Postselection: Finally, the registers are measured and postselection is applied, retaining only those outcomes from the syndrome register that correspond to a correctly decoded state from the error register. This ensures that the final output represents the desired solution.
\end{enumerate}

A concise summary and further insight into the state evolution during the process, refer to Figure~\ref{fig:dqidigram}.


\subsection{Detailed Component Implementation}
\subsubsection{Amplitude Encoding}
A key step in the DQI algorithm is determining the amplitudes \(\{w_k\}\) and then preparing an amplitude-encoded state using these weights. We first compute the weight coefficients by determining the principal eigenvector of a specific \((\ell+1)\times(\ell+1)\) symmetric tridiagonal matrix \(A^{(m,d,\ell)}\). These coefficients are used to construct the state
\begin{equation}
|\psi_{\text{UAE}}\rangle \;=\; \sum_{k=0}^{\ell} w_k \,\lvert k \rangle,
\label{eq:psi_UAE}
\end{equation}

with \(\sum_{k=0}^{\ell} |w_k|^2 = 1\).

Following Jordan et al.~\cite{jordan-dqi}, the matrix \(A^{(m,d,\ell)}\) is given by:
\begin{equation}
A^{(m,d,\ell)} \;=\;
\begin{pmatrix}
0      & a_1   & 0      & \cdots & 0 \\
a_1    & d     & a_2    & \cdots & 0 \\
0      & a_2   & 2d     & \cdots & 0 \\
\vdots &\vdots &\vdots  & \ddots & a_\ell \\
0      & 0     & 0      & a_\ell & \ell d
\end{pmatrix},
\end{equation}
with the off-diagonal elements and the diagonal scaling parameter defined as
\[
a_k = \sqrt{k\,(m-k+1)} \quad\text{and}\quad d = \frac{p - 2r}{\sqrt{r\,(p-r)}}.
\]
Here, \(m\) is the number of constraints, \(p\) is the prime that defines \(\mathbb{F}_p\) (with \(p=2\) in the case of Max-XORSAT), and \(r\) indicates the number of inputs that yield \(+1\) for each constraint (typically, \(r=1\) for Max-XORSAT). The principal eigenvector of \(A^{(m,d,\ell)}\) yields the desired coefficients \(w_k\).

Once these coefficients are determined classically, we employ a Unary Amplitude Encoding (UAE) procedure that's a quantum fan-out operation \cite{ref11} to create a unary state. Specifically, starting from the error registers initialized in the state \(\ket{0}^{\otimes m}\), we apply a sequence of single-qubit \(R_y\) rotations and controlled \(R_y\) (CR\(_y\)) gates. This sequence effectively fans out the amplitude information, thereby distributing the computed coefficients across multiple qubits and preparing the unary state as specified in~\eqref{eq:psi_UAE}.

\subsubsection{Dicke State Preparation}

\begin{figure*}[htbp]
\centering
\scalebox{.6}{
\begin{quantikz}[column sep=0.5cm]
\lstick[wires=8]{$m$}
 & \gate[wires=8]{U_{m,\ell}} & \qw \\
 & \ghost{U_{m,\ell}} & \qw \\
 & \ghost{U_{m,\ell}} & \qw \\
 & \ghost{U_{m,\ell}} & \qw \\
 & \ghost{U_{m,\ell}} & \qw \\
 & \ghost{U_{m,\ell}} & \qw \\
 & \ghost{U_{m,\ell}} & \qw \\
 & \ghost{U_{m,\ell}} & \qw
\end{quantikz}
\quad = \quad
\begin{quantikz}[row sep=.01cm,column sep=0.25cm]
 & \qwbundle{1} & \gate[wires=2][50pt][20pt]{\mathrm{SCS}_{m,\ell}} & \qw & \qw & \qw &  &  & \qw & \qw & \qw & \qw\\
 & \qwbundle{\ell} & \qw  & \gate[wires=2][50pt][20pt]{\mathrm{SCS}_{m-1,\ell}} & \qw & \qw &  &  & \qw & \qw & \qw & \qw\\
 & \qwbundle{1} & \qw & \qw & \gate[wires=2][50pt][20pt]{\mathrm{SCS}_{m-2,\ell}} & \qw &  &  & \qw & \qw & \qw & \qw\\
 & \qwbundle{\ell} & \qw & \qw & \qw & \qw &  &  & \qw & \qw & \qw & \qw \\
 &  &  &  &  &  & \ddots &  & & & & \\
 & \qw & \qw & \qw & \qw & \qw &  & & \gate[wires=6][20pt][30pt]{\mathrm{SCS}_{\ell+1,\ell}} & \qw \slice[style=black]{} & \qw & \qw & \qw & \qw &  & & \qw & \qw & \qw & \qw\\
 & \qw & \qw & \qw & \qw & \qw &  & & \qw & \qw  & \qw & \gate[wires=5][50pt][20pt]{\mathrm{SCS}_{\ell,\ell-1}} & \qw & \qw &  &  & \qw & \qw & \qw & \qw\\
 & \qw & \qw & \qw & \qw & \qw & &  & \qw & \qw & \qw & \qw & \gate[wires=4][50pt][20pt]{\mathrm{SCS}_{\ell-1,\ell-2}} & \qw & \dots &  & \qw & \qw & \qw & \qw\\
 & \qw & \qw & \qw & \qw & \qw & &  & \qw & \qw & \qw & \qw & \qw & \qw &  &  & \qw & \gate[wires=3][20pt][30pt]{\mathrm{SCS}_{3,2}} & \qw & \qw\\
 & \qw & \qw & \qw & \qw & \qw & &  & \qw & \qw & \qw & \qw & \qw & \qw &  &  & \qw & \qw & \gate[wires=2][50pt][20pt]{\mathrm{SCS}_{2,1}} & \qw\\
 & \qw & \qw & \qw & \qw & \qw & &  & \qw & \qw & \qw & \qw & \qw & \qw &  &  & \qw & \qw & \qw & \qw
\end{quantikz}
}
\caption{Inductive construction of \(U_{m,\ell}\) from a sequence of \(\mathrm{SCS}_{m,\ell}\)}
\label{fig:scs-nk-fig}
\end{figure*}
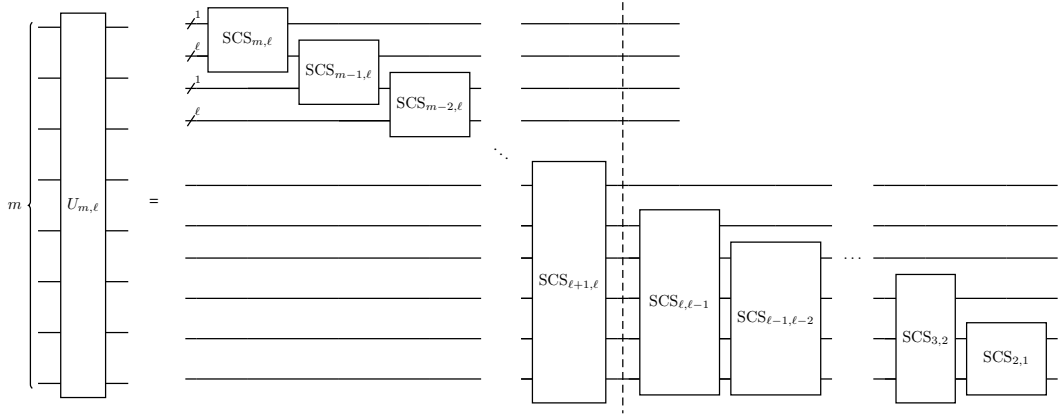
To convert the unary state produced by UAE into the desired Dicke state, we apply a deterministic unitary \(U_{m,\ell}\). This unitary is implemented as a cascade of  Split and Cyclic Shift (SCS) unitaries, denoted \(\mathrm{SCS}_{m,\ell}\), which collectively map the unary encoding onto the Dicke state with \(\ell\) excitations. Following the construction outlined by Bärtschi and Eidenbenz~\cite{ref2}, the SCS unitary is decomposed into elementary operations: the two-qubit version, \(\mathrm{SCS}_2\), and the three-qubit version, \(\mathrm{SCS}_3\), corresponding to building block gates (i) and (ii) in their work.

Each \(\mathrm{SCS}_{m,\ell}\) gate is implemented by applying \(\ell-1\) copies of the \(\mathrm{SCS}_3\) followed by one  \(\mathrm{SCS}_2\). The unitary \(U_{m,\ell}\) is composed of these \(\mathrm{SCS}_{m,\ell}\) in two stages. The first stage applies SCS on overlapping qubit subsets, descending from \(m\) to \(\ell+1\), and the second stage further refines the Dicke state by applying additional SCS descending from \(\ell\) to \(2\). This systematic construction ensures a deterministic Dicke state preparation without postselection.
The inductive composition of \(U_{m,\ell}\) from \(\mathrm{SCS}_{m,\ell}\) unitary is illustrated explicitly in Figure~\ref{fig:scs-nk-fig}.

\subsubsection{Phase Encoding} 
For each \( j \) with \( v_j = 1 \), a Pauli-\( Z \) gate is applied on the \( j \)-th qubit of \(Y\) to encode phase information, transforming the quantum state as follows:
\[
\ket{y} \mapsto (-1)^{v\cdot y}\ket{y}.
\]
\subsubsection{Constraint Encoding}  For each entry \( B^{T}_{ij} = 1 \), apply a CNOT gate with control on the \( j \)-th qubit of the error register \( y \) and target on the \( i \)-th qubit of the syndrome register \( S \). This operation maps the error state \(\ket{y}\) into the syndrome register, yielding the state \(|B^T y\rangle\).

\subsubsection{Decoding}
The decoding stage employs a reversible implementation of Gauss-Jordan elimination to uncompute the error register and isolate the syndrome information. In our design, we compare two methods. Here we elaborate on each approach.

\begin{enumerate}[label=\alph*), leftmargin=3em]
  \item Gauss–Jordan elimination:  
  In this method, the decoding circuit reversibly transforms the augmented matrix \([B^T \mid y]\) into its reduced row‐echelon form, thereby isolating the syndrome state \(|B^T y\rangle\). This approach uses SWAP and CNOT operations as described in Algorithm~\ref{alg:gaussjordan}.
  \item Lookup Table:  
In the lookup table approach, every possible error pattern and its corresponding syndrome is precomputed classically. The quantum decoding step applies gates conditionally based on the syndrome state. Specifically, for each syndrome–error pair, the syndrome register (control) is prepared to match the precomputed syndrome state by applying \(X\) gates to syndrome qubits that must be controlled on \(\ket{0}\). This preparation step is done twice (before and after the controlled operation), requiring at most \(2m\) \(X\) gates per syndrome pattern. Subsequently, multi-controlled \(X\) gates (\(C^n (X)\)) are applied with controls on the syndrome register and targets on the error register. Summing over all error patterns of weight up to \(\ell\), the total gate count can grow significantly \cite{ref10}.
\end{enumerate}

\subsubsection{Measurement and Postselection}
Finally, all registers are measured. Due to potential imperfections in the decoding circuit, there is a possibility of obtaining erroneous states in the error register \(Y\). Therefore, we apply postselection by retaining only measurement outcomes where the error register \(Y\) is found in the state \(\ket{0}^{\otimes m}\). Under this condition, the syndrome register \(S\) reliably encodes the correctly decoded solution.


\section{Circuit Analysis}
\begin{table*}[htbp]
    \caption{Gate Count and Depth Complexity for DQI Circuit Components}
    \label{tab:complexity_simplified}
    \centering
    \resizebox{\textwidth}{!}{%
      \begin{tabular}{l|c|c|c|c|c|c|c}
      \hline
      \textbf{Gate} 
      & \thead{Unary Amplitude \\ Encoding}
        & \thead{Dicke State Preparation} 
        & \thead{Phase \\ Encoding} 
        & \thead{Constraint \\ Encoding} 
        & \thead{Gauss-Jordan \\ Elimination} 
        & \textbf{Lookup Table} 
        & \thead{Hadamard \\ Transform} \\
      \hline
      \textbf{X}           
        & \(0\) 
        & \(0\) 
        & \(0\) 
        & \(0\) 
        & \(0\) 
        & \(\leq 2m\,\displaystyle \sum\nolimits_{r=0}^{\ell} \binom{n}{r}\) 
        & \(0\) \\
      \textbf{Z}           
        & \(0\) 
        & \(0\) 
        & \(\leq m\) 
        & \(0\) 
        & \(0\) 
        & \(0\) 
        & \(0\) \\
    \textbf{Hadamard}
        & \(0\)
        & \(0\)
        & \(0\)
        & \(0\)
        & \(0\)
        & \(0\)
        & \(n\) \\
      \(\mathbf{R_y}\) 
        & \(2\ell+1\) 
        & \(6m\ell - 4m - 3\ell^2 - 3\ell + 4\) 
        & \(0\) 
        & \(0\) 
        & \(0\) 
        & \(0\) 
        & \(0\) \\
      \textbf{CNOT}        
        & \(2\ell\) 
        & \(10m\ell - 6m - 5\ell^2 - 5\ell + 6\) 
        & \(0\) 
        & \(\leq mn\) 
        & \(\leq mn\) 
        & \(0\) 
        & \(0\) \\
      \(\mathbf{C^n (X)}\)        
        & \(0\) 
        & \(0\) 
        & \(0\) 
        & \(0\) 
        & \(0\) 
        & \(\leq m\,\displaystyle \sum\nolimits_{r=0}^{\ell}\binom{n}{r}\) 
        & \(0\) \\
      \textbf{SWAP}       
        & \(0\) 
        & \(0\) 
        & \(0\) 
        & \(0\) 
        & \(\leq n\) 
        & \(0\) 
        & \(0\) \\
      \hline
      \textbf{Depth}       
        & \(3\ell+1\) 
        & \(\leq (m-1)(16\ell-10)\) 
        & \(1\) 
        & \(\leq mn\)  
        & \(\leq n(m+1)\) 
        & \(\leq (m+2)\displaystyle \sum\nolimits_{r=0}^{\ell}\binom{n}{r}\) 
        & \(1\) \\
      \textbf{Qubits}      
        & \(2^{\lceil \log_2 (\ell+1) \rceil}\) 
        & \(m\) 
        & \(m\) 
        & \(m+n\) 
        & \(m+n\) 
        & \(m+n\) 
        & \(n\) \\
      \hline
      \end{tabular}%
    }
\end{table*}

In our analysis, the problem is defined by a constraint matrix \(B \in \{0,1\}^{m \times n}\), where \(m\) is the number of constraints and \(n\) is the number of variables. Accordingly, the syndrome register is of size \(m\) and the error register is of size \(n\). Moreover, the degree \(\ell\) of the polynomial representing the final solution state is chosen so that its optimal value corresponds to the maximum number of errors possible in \(B\).

\subsection{Unary Amplitude Encoding (UAE)}
In this stage, the goal is to encode the weight vector \(\{w_k\}_{k=0}^{\ell}\) as well as its sum into a unary state. Specifically, the unary state is intended to represent the amplitudes corresponding to each weight \(w_k\) (with an additional state for the zero weight), i.e., 
\[
\sum_{k=0}^{\ell} w_k \,|k\rangle.
\]
To achieve this encoding, controlled \(R_y\) rotations are applied—one for each level from \(0\) to \(\ell\) (yielding \(\ell+1\) levels in total). Each controlled \(R_y\) gate is decomposed into 2 \(R_y\) rotations and 2 CNOT gates, which explains the gate counts presented in the table. For the circuit depth, since the first \(R_y\) rotation in each controlled gate sets the primary depth, we count 3 layers per controlled gate plus an additional 1 layer overall, resulting in a depth of \(3\ell+1\).

The number of qubits required for the UAE stage is determined by the size of the register needed to represent \(\ell+1\) distinct states. Specifically, we choose the smallest integer \(\alpha\) such that \(2^\alpha > \ell\), so that the unary state is encoded in a register of size
\[
2^{\lceil \log_2 (\ell+1) \rceil}.
\]

\subsection{Dicke State Preparation}
The Dicke state preparation unitary \(U_{m,\ell}\) is constructed from multiple \(\mathrm{SCS}_{m,\ell}\) unitaries. Each \(\mathrm{SCS}_{m,\ell}\) requires a specific number of elementary gates. \(\mathrm{SCS}_2\) consists of \(4\) CNOT gates and \(2\) \(R_y\) gates. \(\mathrm{SCS}_3\) consists of \(2\) CNOT gates and one controlled-controlled \(R_y\) gate (\(\mathrm{CCR}_y\)), where each \(\mathrm{CCR}_y\) decomposes into \(2\) additional CNOT gates and \(3\) single-qubit \(R_y\) gates. Thus, each \(\mathrm{SCS}_3\) has a total of \(4\) CNOT gates and \(3\) \(R_y\) gates.

More generally, an \(\mathrm{SCS}_{m,\ell}\) is composed of \(\ell-1\) copies of \(\mathrm{SCS}_3\) and one copy of \(\mathrm{SCS}_2\), resulting in:
\begin{equation}
10(\ell-1)+4 = 10\ell - 6 \quad \text{CNOT gates,}
\end{equation}
and
\begin{equation}
6(\ell-1)+2 = 6\ell - 4 \quad R_y \text{ gates.}
\end{equation}

The complete unitary \(U_{m,\ell}\) applies these gates in two distinct phases:

- \emph{Phase one:} Apply \(\mathrm{SCS}_{l,\ell}\) for \(l = m, m-1, \dots, \ell+1\), yielding a total of \(m-\ell\) gates, each using \(10\ell - 6\) CNOT gates and \(6\ell - 4\) \(R_y\) gates. 

- \emph{Phase two:} Apply \(\mathrm{SCS}_{l,l-1}\) for \(l = \ell, \ell-1, \dots, 2\), resulting in \(\ell-1\) gates, each gate using \(10(l-1)-6 = 10l - 16\) CNOT gates and \(6(l-1)-4 = 6l - 10\) \(R_y\) gates.

Thus, the total gate counts are:\\
- CNOT gates:
\begin{equation}
( m-\ell)(10\ell-6) + \sum_{l=2}^{\ell}(10l-16) = 10m\ell - 6m - 5\ell^2 - 5\ell + 6,
\end{equation}
\\
- \(R_y\) gates:
\begin{equation}
( m-\ell)(6\ell-4) + \sum_{l=2}^{\ell}(6l-10) = 6m\ell - 4m - 3\ell^2 - 3\ell + 4.
\end{equation}

The depth of \(U_{m,\ell}\) is estimated by considering that each of the \(m-1\) layers consists of an \(\mathrm{SCS}_{m,\ell}\) with depth approximately equal to the sum of depths from its constituent gates (\(10\ell - 6\) for CNOT and \(6\ell - 4\) for \(R_y\)), leading to a depth bound of:
\begin{equation}
(m-1)(16\ell - 10).
\end{equation}

Since the Dicke state is prepared on the \(m\)-qubit register associated with the constraints, the Dicke state preparation stage utilizes \(m\) qubits in total. All gate-count and depth complexities are summarized in Table~\ref{tab:complexity_simplified}.

\subsection{Phase Encoding}
Phase information corresponding to the problem constraints is imprinted on the \(m\)-qubit syndrome register via Pauli-\(Z\) gates. In particular, for each entry in the constraint vector \(v\) (with at most \(m\) ones), a \(Z\) gate is conditionally applied to the corresponding qubit. Since the number of ones in \(v\) can be at most \(m\), up to \(m\) \(Z\) gates are applied. Moreover, because these \(Z\) gates act on distinct qubits, they can be executed in parallel, so the phase encoding stage operates at constant depth.

\subsection{Constraint Encoding}
The syndrome is generated by applying CNOT operations that implement the transformation \(|B^T y\rangle\). Specifically, the error register \(y\) (consisting of \(m\) qubits) is used as the control, and the transformation \(B^T\) is applied to the syndrome register (consisting of \(n\) qubits) as the target, thereby preparing the state \(|B^T y\rangle\) on the syndrome register. In the worst-case scenario, up to \(mn\) CNOT gates are required, with the circuit depth bounded by \(\leq mn\) layers when applied sequentially. Overall, this encoding stage uses a total of \(m+n\) qubits.

\subsection{Decoding}
Decoding the syndrome is a critical step in the DQI algorithm. In our design, two methods are compared. Here, we elaborate on each approach.

\begin{enumerate}
  \item Gauss–Jordan elimination:\\  
  In this method, the decoding circuit reversibly transforms the augmented matrix 
  \(\bigl[B^T \mid s\bigr]\) (with \(s = B^T y\)) into 
  \(\bigl[I \mid y\bigr]\) in reduced row-echelon form, thereby computing \(\ket{y}\) from \(\ket{B^T y}\).  
  The procedure operates as follows:
  \begin{enumerate}
    \item Pivot Selection and Row Swapping: For each column \(j = 1, \ldots, n\), the circuit scans to identify a pivot (an entry equal to 1). If the pivot is not in the current row, a SWAP gate is applied to exchange rows and correctly position the pivot. This procedure uses up to \(n\) SWAP gates.
    \item Row Elimination: Once the pivot is positioned, CNOT gates are applied from the pivot row (control) to every other row that has a 1 in the pivot column (target), zeroing out the non-pivot entries. This step requires up to \(m-1\) CNOT gates per column, totaling up to \(n(m-1)\) CNOT gates.
    \item Error Uncomputation: Once the syndrome register holds \(\ket{y}\), apply CNOT from each syndrome qubit (control) to the corresponding error qubit (target) to uncompute the original error register. This step requires up to \(n\) additional CNOT gates.
  \end{enumerate}
  Overall, the Gauss–Jordan elimination approach employs up to
\begin{equation}
n(m-1) + n = nm
\end{equation}
CNOT gates and up to \(n\) SWAP gates, leading to a circuit depth bounded by \(mn+n = n(m+1)\) layers. This procedure is executed on a combined register of \(m+n\) qubits, where \(m\) qubits form the error register and \(n\) qubits form the syndrome register.

  \item Lookup Table Method:\\  
In the lookup table approach, every possible error pattern and its corresponding syndrome is precomputed classically. The quantum decoding step applies gates conditionally based on the syndrome state. Specifically, for each syndrome–error pair, the syndrome register (control) is prepared to match the precomputed syndrome state by applying \(X\) gates to syndrome qubits that must be controlled on \(\ket{0}\). This preparation step is done twice (before and after the controlled operation), requiring at most \(2m\) \(X\) gates per syndrome pattern. Subsequently, multi-controlled \(X\) gates (\(C^n(X)\)) are applied from the syndrome register (control) to each corresponding qubit in the error register (target) that must be flipped according to the precomputed error pattern. Since each \(C^n(X)\) gate flips exactly one qubit in the error register, the circuit applies up to \(m\) of these gates per syndrome-error pair.

Summing over all error patterns of weight up to \(\ell\), the total number of \(X\) gates used for control preparation is at most
\begin{equation}
2m\sum_{r=0}^{\ell}\binom{n}{r},
\end{equation}
and the total number of \(C^n(X)\) gates applied is at most
\begin{equation}
m\sum_{r=0}^{\ell}\binom{n}{r}.
\end{equation}

The overall depth of the lookup table decoding stage is bounded by
\begin{equation}
(m+2)\sum_{r=0}^{\ell}\binom{n}{r},
\end{equation}
and this method uses a combined register of \(m+n\) qubits.

\end{enumerate}

After applying the decoding unitary to compute \(\ket{y}\) from \(\ket{B^T y}\) on the syndrome register and updating the error register, we uncompute the decoder by applying its inverse, which returns the syndrome register from \(\ket{y}\) back to \(\ket{B^T y}\).  

In summary, while both decoding methods operate on the same \((m+n)\)-qubit register, the Gauss–Jordan elimination approach generally offers significantly lower gate counts and circuit depth compared to the lookup table method.   

\subsection{Hadamard Transform}
After the decoding stage, the Hadamard transform is applied to the \(n\)-qubit syndrome register to convert the quantum state into the computational basis suitable for measurement. Specifically, a single-qubit Hadamard gate is applied independently to each of the \(n\) qubits. This step thus uses exactly \(n\) Hadamard gates and operates at depth \(1\), employing \(n\) qubits.


\section{Results and Discussion}
\label{sec:results}

In this work, we address the MaxCut problem by formulating it as a special instance of the Max-XORSAT problem. In the standard Max-XORSAT formulation, one is given a collection of XOR constraints over a finite field. For our purposes, these constraints are defined over \(\mathbb{F}_2\) (i.e., \(p=2\)) and are satisfied when the corresponding linear equation evaluates to 1, so that the cardinality parameter is \(r=1\).  

\subsection{MaxCut as a Special Instance of Max-XORSAT}

The MaxCut problem is a well-known combinatorial optimization task: given a graph \(G=(V,E)\) (or weighted graph), the objective is to partition the vertex set \(V\) into two disjoint subsets such that the number (or total weight) of edges crossing the partition is maximized. By assigning a binary variable \(x_i \in \{0,1\}\) to each vertex \(i\), each edge \((i,j)\in E\) defines an XOR constraint
\[
x_i \oplus x_j = 1,
\]
which is equivalent to requiring that the endpoints of the edge lie in different subsets. In the Max-XORSAT framework, each constraint contributes \(+1\) if satisfied and \(-1\) if unsatisfied.

For MaxCut, each constraint enforces that the two endpoints of an edge lie in different subsets. Therefore, we set \(v_i = 1\) for every constraint leading to \(v\) being a vector with all elements equal to 1. In this formulation, each constraint contributes \(\pm1\) to \(f(x)\), so the difference between satisfied and unsatisfied cases is always 2; accordingly, the maximum number of bit-flip “errors” to distinguish is 2, and thus a degree-2 polynomial (\(\ell=2\)) suffices to discriminate and amplify the near-optimal solutions.

\subsection{Simulation Results and Resource Metrics}

With \(\ell=2\) fixed, the Unary Amplitude Encoding stage requires only three amplitude levels (with a register size of \(2^{\lceil \log_2 3\rceil} = 4\)). The Dicke state preparation, phase encoding, constraint encoding, and decoding stages are optimized accordingly. Table~\ref{tab:resource_comparison} summarizes the resource metrics in gate count, depth, and qubit usage obtained from our implementation when transpiled into our chosen universal gate set, namely \(\{Z,\;\mathrm{CNOT},\;R_x,\;R_z,\;R_y,\;\mathrm{SWAP}\}\), using the Qiskit Transpiler~\cite{ref14} rather than by directly counting the quantum runtime as in \cite{ref5}, since our experiments are performed in simulation.These basis gates align with the native instruction sets of leading quantum hardware—Pauli-\(Z\) rotations can be implemented virtually with zero error and duration, while CNOT and SWAP map directly to the two-qubit entangling operations on superconducting devices

\begin{table*}[!ht]
  \caption{Resource Metrics for DQI Circuit Components (After Transpilation into \(\{Z,\,\mathrm{CNOT},\,R_x,\,R_z,\,R_y,\,\mathrm{SWAP}\}\))}
  \label{tab:resource_comparison}
  \centering
  \begin{tabular}{l|ccc|ccc|ccc}
    \hline
    \textbf{Subroutine} & \multicolumn{3}{c|}{\textbf{Gate Count}} & \multicolumn{3}{c|}{\textbf{Depth}} & \multicolumn{3}{c}{\textbf{Qubits}} \\
    \textbf{instance size} & \textbf{m=5} & \textbf{m=10} & \textbf{m=15} & \textbf{m=5} & \textbf{m=10} & \textbf{m=15} & \textbf{m=5} & \textbf{m=10} & \textbf{m=15} \\
    \hline
    Unary Amplitude Encoding & 14 & 14 & 14 & 6 & 6 & 6 & 5 & 10 & 15 \\
    Dicke State Preparation & 110 & 285 & 452 & 77 & 196 & 320 & 5 & 10 & 15 \\
    Phase Encoding & 5 & 10 & 15 & 1 & 1 & 1 & 5 & 10 & 15 \\
    Constraint Encoding & 10 & 20 & 30 & 6 & 5 & 8 & 10 & 20 & 30 \\
    Decoding (GJE) & 12 & 17 & 46 & 10 & 17 & 39 & 10 & 20 & 30 \\
    Decoding (Lookup Table) & 1885 & 17112 & 67027 & 1655 & 15778 & 62547 & 10 & 20 & 30 \\
    Hadamard Transform & 10 & 20 & 30 & 2 & 2 & 2 & 5 & 10 & 15 \\
    \hline
  \end{tabular}
\end{table*}

From Table~\ref{tab:resource_comparison}, we can see that the Dicke State Preparation subroutine has the largest gate count and depth among the circuit components for these smaller instances. However, because \(\ell=2\) remains constant for MaxCut, the overall Dicke state preparation cost grows only \emph{linearly} with \(m\). By contrast, the Constraint Encoding and Decoding steps scale \emph{polynomially} in \(m\), in line with their theoretical \(O(m^2)\) behavior (especially for the decoding).

Moreover, comparing the “Decoding (GJE)” rows with the “Decoding (Lookup Table)” rows, we observe that the Gauss–Jordan Elimination (GJE) approach requires drastically fewer gates and achieves shallower circuit depths than the lookup table method. This efficiency gap grows with \(m\); at \(m=15\), GJE uses fewer than 50 gates, whereas the lookup table decoding needs tens of thousands.

\begin{figure}[ht]
    \centering
    \includegraphics[width=\linewidth]{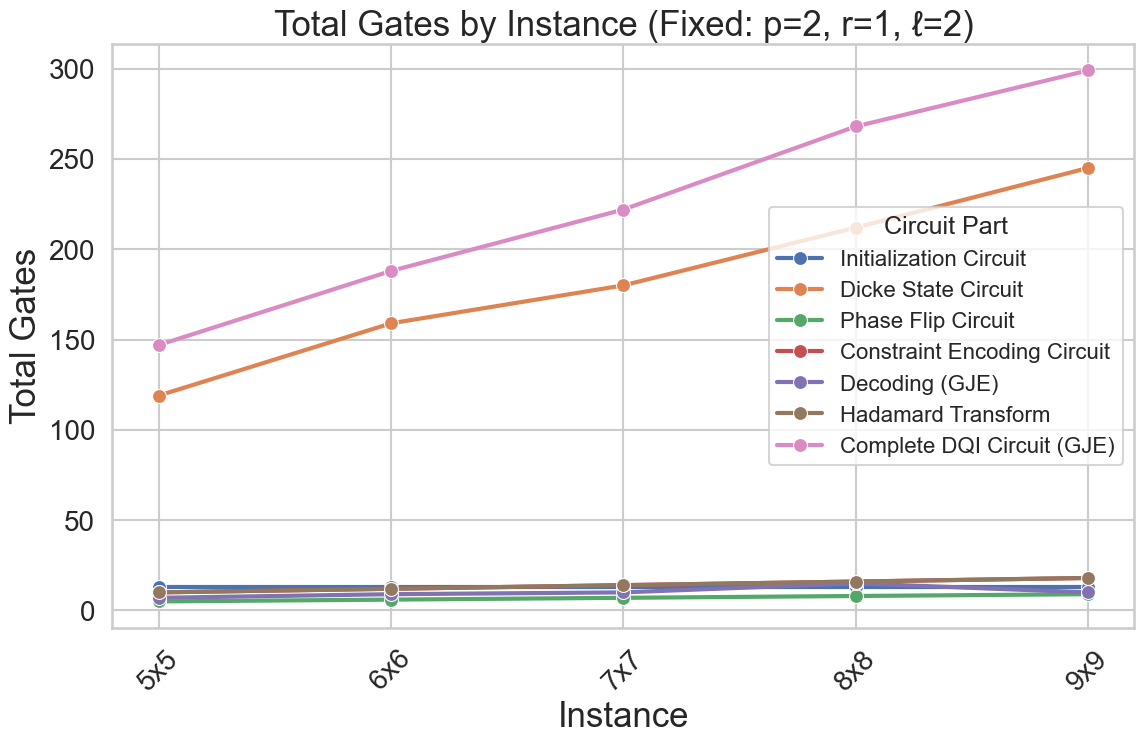}
    \caption{Total gate count for each circuit component versus instance size, with \(p=2\), \(r=1\), and \(\ell=2\).}
    \label{fig:total_gates}
\end{figure}

\begin{figure}[ht]
    \centering
    \includegraphics[width=\linewidth]{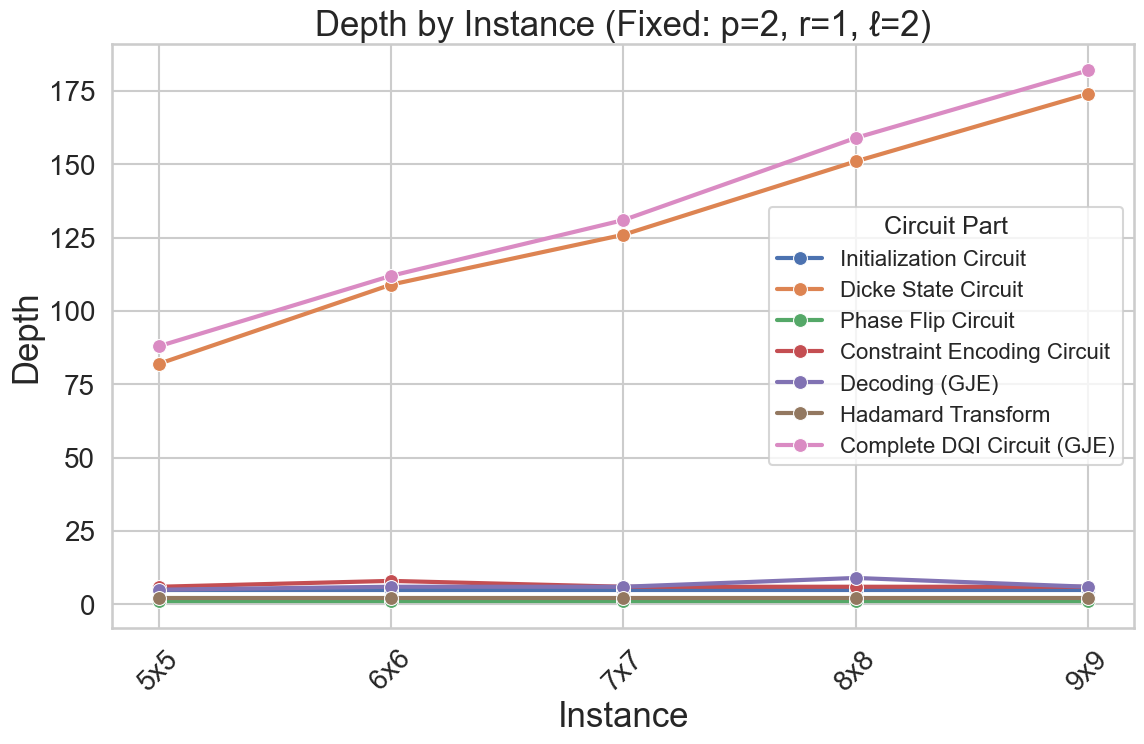}
    \caption{Circuit depth versus instance size, with \(p=2\), \(r=1\), and \(\ell=2\).}
    \label{fig:depth}
\end{figure}

Figures~\ref{fig:total_gates} and \ref{fig:depth} show the total gate count and circuit depth, respectively, as the instance size \(m\) increases.  

Notably, Dicke State Preparation visibly dominates the smaller-scale portion of the plots. However, as \(m\) increases, the polynomial overhead from Constraint Encoding and Decoding eventually becomes the main bottleneck. Overall, these simulations confirm that while Dicke State Preparation is a major contributor at small \(m\), the choice of \(\ell=2\) keeps its scaling linear. Meanwhile, the decoding approach based on GJE provides a practical advantage over the lookup table method, maintaining much lower gate counts and circuit depths in practice.

\subsection{Example Result for a 6-Bit Instance}
\begin{figure*}[!t]
    \centering
    \includegraphics[width=0.95\linewidth]{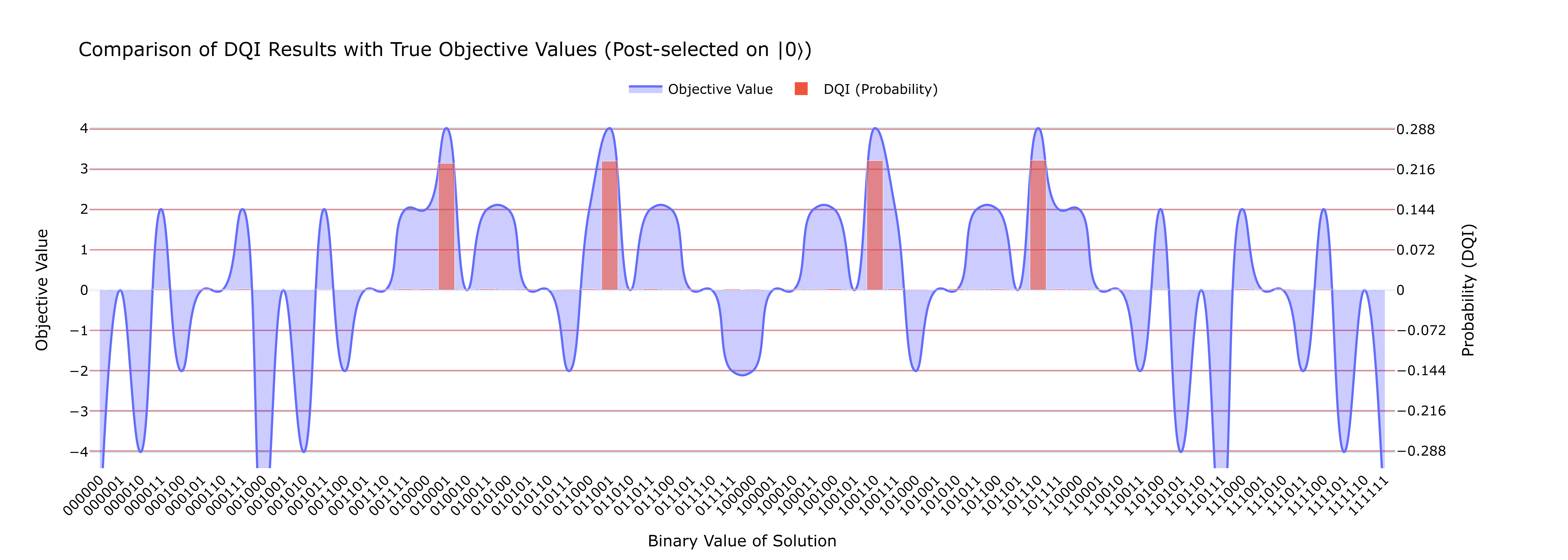}
    \caption{Result plot for the 6-bit instance. The \(x\)-axis lists all possible 6-bit states. The blue line (left \(y\)-axis) shows the classical objective value for each state, while the red bars (right \(y\)-axis) display the measured probability (post-selected on \(\ket{0}\)). The coincidence of the objective function peaks with large measurement probabilities shows clearly that DQI finds optimal (or near-optimal) peaks.}
    \label{fig:dqi_output_6bit}
\end{figure*}

For further illustration, we consider a 6-bit instance. The input for this instance is defined by the following matrix \(B\) and vector \(v\):

\vspace{0.5em}
\noindent
\textbf{Matrix (\(B\)):}
\[
B =
\begin{pmatrix}
1 & 0 & 0 & 0 & 0 & 1 \\
1 & 1 & 0 & 0 & 0 & 0 \\
0 & 1 & 0 & 0 & 1 & 0 \\
0 & 1 & 0 & 1 & 0 & 0 \\
0 & 1 & 0 & 0 & 0 & 1 \\
0 & 0 & 0 & 1 & 0 & 1
\end{pmatrix}
\]
\textbf{Vector (\(v\)):}  
For the MaxCut problem, \(v\) is always a vector of all 1's since every constraint is expected to be satisfied to yield a maximum cut, $v=(1\ 1 \ 1 \ 1 \ 1 \ 1 )$.

\vspace{0.5em}
After executing the DQI circuit and post-selecting on the error register being \(\ket{0}^{\otimes 6}\), we obtain a probability distribution over the \(2^6 = 64\) possible 6-bit solutions. Figure~\ref{fig:dqi_output_6bit} shows the result plot where:
\begin{enumerate}
  \item The \(x\)-axis lists all possible 6-bit solution states (from \(\ket{000000}\) to \(\ket{111111}\)).
  \item The left \(y\)-axis (blue curve) represents the classical objective function values.
  \item The right \(y\)-axis (red bars) indicates the measured probability of each solution.
\end{enumerate}
In the plot, higher probabilities (red bars) tend to coincide with higher objective values (blue line), demonstrating that the DQI algorithm successfully biases the measurement outcomes toward optimal or near-optimal solutions.

\subsection{Discussion}

Our implementation, publicly available at \url{https://github.com/BankNatchapol/DQI-Circuit}, is general and supports arbitrary matrix dimensions (both square and non-square) and handles general cases of the Max-XORSAT problem.

Due to limitations in our simulation environment, we were only able to simulate systems of up to 30 qubits (corresponding to instance sizes of \(m=15\)). As a consequence, we do not yet observe a saturating satisfiability fraction that would fully demonstrate the theoretical advantages of the approach at larger scales.

In the original DQI formulation~\cite{jordan-dqi}, the authors observed that an imperfect decoder leaves misidentified errors that are typically filtered out by post-selecting on the error register being \(\ket{0}^{\otimes m}\). However, because different decoding algorithms vary in their capacity, the structure of these misidentified errors can differ substantially. In our experiments, we found that combining outcomes from imperfect decoding (with nonzero syndrome registers) and correct decoding sometimes boosts the target state probability. This suggests that misidentified errors may carry useful information about the overall error distribution, and that a more refined post-selection strategy—rather than a strict cutoff—could enhance performance. Further investigation is required to determine under which conditions this combined strategy is most effective.

Finally, although our study focuses on quantum circuit complexity, we have not analyzed the classical overhead for parameter optimization and post-processing. Assessing this classical component is essential for evaluating the algorithm's end-to-end efficiency.

\subsection{Conclusion and Future Work}

This paper presented a detailed quantum circuit design for the Decoded Quantum Interferometry (DQI) algorithm applied to solving a MaxCut problem instance. Our design integrates Unary Amplitude Encoding, deterministic Dicke state preparation via SCS gates, phase encoding, constraint encoding, a Gauss--Jordan elimination-based decoding circuit, and Hadamard transform. Theoretical and simulated results indicate that the overall gate count and circuit depth scale roughly as \(O(m^2)\), while the circuit requires approximately \(2m\) qubits.

Future work will focus on:
\begin{enumerate}
    \item Expanding the algorithmic framework to more general problem types such as Max-LINSAT and other combinatorial optimization problems.
    \item Conducting a more comprehensive analysis of both the classical and quantum components of the algorithm, including resource overheads in parameter pre-processing and measurement post-processing.
    \item Scaling the instance sizes further to demonstrate the algorithm's advantages at larger scales.
    \item Experimentally validating the circuit on near-term quantum hardware to estimate noise thresholds and assess the robustness of the algorithm under realistic conditions.
    \item Developing more efficient implementations for each circuit component, particularly optimizing the decoding stage. In this regard, recent work~\cite{ref12} presents an improved decoding algorithm that may help reduce gate counts and circuit depth for the decoding step.
    \item Investigating alternative approaches for Dicke state preparation based on the methods described inBärtschi and Eidenbenz~\cite{ref13}, Wang, Cong, and De Micheli~\cite{ref23}, which may further improve the efficiency of state preparation.
\end{enumerate}

These efforts will contribute to a deeper understanding of the practical advantages and limitations of the DQI algorithm for combinatorial optimization problems.

\section*{Acknowledgement}
NP thanks Stephen P. Jordan for helpful discussion. We acknowledge the use of ChatGPT to assist in improving the clarity and language of this manuscript.

\section*{}
\nocite{*}
\bibliographystyle{IEEEtran.bst}
\bibliography{IEEEabrv, references}

\end{document}